\documentclass[12pt]{article}
\usepackage{amsmath,amssymb}
\usepackage[mathscr]{euscript}
\usepackage{graphicx}
\usepackage[all]{xypic}

\newtheorem{prop}{Proposition}[section]

\newtheorem{lemm}{Lemme}[section]

\newcommand{\iN}{\hbox{ {\leaders\hrule\hskip.2cm}{\vrule height
      .22cm} }}

\newcommand{\R}{\mathbb{R}}

\renewcommand{\H}{\mathbb{H}}
\newcommand{\N}{\mathbb{N}}

\title{Manifolds obtained by soldering together points, lines, etc.}

\date{}

\author{Fr{\'e}d{\'e}ric \textsc{H{\'e}lein}\footnote{Institut de Math{\'e}matiques de Jussieu, UMR CNRS 7586, Universit{\'e} Paris Diderot--Paris 7,
Case 7012, b{\^a}timent Chevaleret, 75205 Paris cedex 13, France, \textsf{helein@math.jussieu.fr}}
}

\begin{document}
\maketitle

\thispagestyle{empty}

\section{The solder form}

\subsection{A physical motivation : ideas from the general\\ relativity}\label{motivation}

One fundamental concept in the theory of general relativity is that the gravitation should not be considered as a force which takes place in some \emph{a priori} given space-time but that, instead, the space-time is built out of the gravitational field. A mathematical achievement of this idea is the Einstein equation for a (pseudo-)metric tensor $g_{ij}$ on a manifold $\mathcal{M}$ (the space-time):
\begin{equation}\label{einstein}
R_{ij} - {1\over 2}Rg_{ij} = T_{ij},
\end{equation}
where $R_{ij}$ is the Ricci curvature tensor of the metric and $R$ is its scalar curvature. This equation tells us how the metric tensor $g_{ij}$ (i.e. the field mediating the gravitation) is obeying to constraints imposed by the energy-momentum tensor $T_{ij}$, which encodes informations about the distribution of energy and momentum in spacetime. For simplicity we will assume in the following that $T_{ij} = 0$, i.e. we consider the Einstein equation in the vacuum. Note that if we work with equation (\ref{einstein}) in the same way as for any other partial differential equation, then it is very easy to forget about the original intuition of Einstein (which is that the space-time is \emph{built out} of the gravitational field $g_{ij}$) and erroneously to implicitely assume that we are \emph{given} some manifold $\mathcal{M}$, and we are looking for an unknown field $g_{ij}$ satisfying (\ref{einstein}). 

We want to take this physical intuition seriously into account and to look for some mathematical framework which would help us to keep in mind that the manifold $\mathcal{M}$ and the metric should be built simultaneously when solving equation (\ref{einstein}). From this point of view the only `kinematic' condition which is imposed is that, at each point of the space-time, the tangent space to it is endowed with a metric (which is a Minkowski metric in the physical case of pseudo-Riemannian manifolds and an Euclidean one in the Riemannian analogous problem). Then the field $(g_{ij})$ describes the way these metrics depend on the point in a smooth way and the Einstein equation (\ref{einstein}) is the `dynamical' constraint on $g_{ij}$. So we have to imagine an infinite continuous family of copies of the same Minkowski or Euclidean space and to find a way to sew together these infinitesimal pieces into a manifold, by respecting (\ref{einstein}).

I was looking for a long time at a satisfactory mathematical framework where these intuitive considerations would fit naturally. Recently I became conscious that such a framework exists for several decades and that I was aware of it: it is based  on the use of the \emph{solder form} (\emph{forme de soudure} in french) introduced by Charles Ehresmann, who was himself strongly inspired by the work of Elie Cartan. I was helped in particular by discussions with Daniel Bennequin and Michel Egeileh about geometric formulations of the theory of supergravity, where the solder form is used extensively and by the knowledge of the recent work by Nabil Kahouadji \cite{kahouadji}.

\subsection{The solder form on a vector bundle}

The idea of a solder form on a vector bundle is a refinement of the notion of a moving frame $(e_1,\cdots ,e_m)$ on a manifold of dimension $m$ and of its dual coframe $(\varphi^1,\cdots ,\varphi^m)$. Recall that one characterization of the coframe $(\varphi^1,\cdots ,\varphi^m)$ is that\footnote{We use here the usual convention on repeated indices, so that $e_i\otimes \varphi^i$ means $\sum_{i=1}^me_i\otimes \varphi^i$.} $e_i\otimes \varphi^i$ coincides with the identity automorphism $Id_{T\mathcal{M}}$ of $T\mathcal{M}$. Here $Id_{T\mathcal{M}}$ is the section of the vector bundle of endomorphisms of the tangent bundle $T\mathcal{M}$ whose value at each point $\textsc{m}\in \mathcal{M}$ is the identity map $T_\textsc{m}\mathcal{M}\longrightarrow T_\textsc{m}\mathcal{M}$. Then a metric on $T\mathcal{M}$ is given by claiming that $(e_1,\cdots ,e_m)$ is orthonormal, so that the choice of the metric is encoded in the choice of a moving frame. The idea of a solder form on a vector bundle\footnote{The solder form makes sense in a broader sense on any fibre bundle which admits a structure group and is then an important ingredient in the description of a \emph{Cartan geometry} and a \emph{Cartan connection} introduced by Charles Ehresmann \cite{ehresmann}. However here we only need its vector bundle version.} is a subtle variant, in which we are considering an auxiliary vector bundle $V\mathcal{M}$ which is isomorphic to $T\mathcal{M}$ and hence in particular has the same dimension $m$ as $\mathcal{M}$. Then we replace the identity automorphism $Id_{T\mathcal{M}}:T_\textsc{m}\mathcal{M}\longrightarrow T_\textsc{m}\mathcal{M}$ by an isomorphism $\varphi$ from $T\mathcal{M}$ to $V\mathcal{M}$, i.e. a section of $V\mathcal{M}\otimes_\mathcal{M}T^*\mathcal{M}$, the vector bundle of linear maps from $T\mathcal{M}$ to $V\mathcal{M}$, such that, $\forall\textsc{m}\in \mathcal{M}$, $\varphi_\textsc{m}$ is an isomorphism. If we moreover have fixed a metric $g$ on $V\mathcal{M}$, we automatically get a Riemannian metric $\varphi^*g$ on $\mathcal{M}$ defined by 
\[
\forall \textsc{m}\in \mathcal{M}, \forall \xi,\eta\in T_\textsc{m}\mathcal{M},\quad (\varphi^*g)_\textsc{m}(\xi,\eta):= g_\textsc{m}(\varphi_\textsc{m}(\xi), \varphi_\textsc{m}(\eta)).
\]
Note that $\varphi$ may alternatively be viewed as a 1-form on $\mathcal{M}$ with values in $V\mathcal{M}$. This 1-form is called the \emph{solder form}. This construction is related to the moving frame description since, once we are given a moving frame $(E_1,\cdots ,E_m)$ on $V\mathcal{M}$, we get automatically a moving frame $(e_1,\cdots ,e_m)$ on $T\mathcal{M}$ defined by :
\begin{equation}\label{phie=E}
\forall \textsc{m}\in \mathcal{M}, \forall i= 1,\dots ,m,\quad
\varphi_\textsc{m}(e_i) = E_i.
\end{equation}
A direct consequence is that the coframe $(\varphi^1,\cdots ,\varphi^m)$ dual to $(e_1,\cdots,\linebreak e_m)$ is just composed of the components $\varphi^i$ of $\varphi$ in the basis $(E_1,\cdots ,E_m)$, i.e. such that $\varphi = E_1\varphi^1 + \cdots + E_m\varphi^m$. Lastly if furthermore the frame $(E_1,\cdots ,E_m)$ is orthonormal for $g$, then $(e_1,\cdots ,e_m)$ is orthonormal for the metric $\varphi^*g$. 

We moreover assume that $V\mathcal{M}$ is equipped with a connection $\nabla$ which respects the metric $g$. Note that, in contrast with the Levi-Civita connection on a Riemannian manifold, this connection is not unique a priori. We consider $\varphi^*\nabla$, the pull-back connection of $\nabla$ by $\varphi$, acting on sections of $T\mathcal{M}$, which can be defined as follows: if $(E_1,\cdots ,E_m)$ and $(e_1,\cdots ,e_m)$ are moving frames on $V\mathcal{M}$ and $T\mathcal{M}$ respectively and if they are related by (\ref{phie=E}), then
\[
\nabla E_j = E_i\omega^i_j\quad \Longleftrightarrow \quad (\varphi^*\nabla) e_j = e_i\omega^i_j,
\]
i.e. the connexion forms $\omega^i_j$'s coincide through our choice of moving frames. This connection clearly respects the induced metric $\varphi^*g$ and it is well-known that $\varphi^*\nabla$ coincides with the Levi-Civita connection on $(\mathcal{M},\varphi^*g)$, i.e. is \emph{torsion free}, iff
\begin{equation}\label{torsionfree}
d\varphi^i + \omega^i_j\wedge \varphi^j = 0,
\end{equation}
where the $\omega^i_j$'s are the connection 1-forms. We can write the relation (\ref{torsionfree}) in a shorter form:
\begin{equation}\label{torsionfree1}
d^\nabla \varphi = 0,
\end{equation}
where $d^\nabla$ is the covariant exterior differential, acting from $V\mathcal{M}\otimes_\mathcal{M}\Omega^p(\mathcal{M})$ to $V\mathcal{M}\otimes_\mathcal{M}\Omega^{p+1}(\mathcal{M})$ and is defined by $d^\nabla\left(E_i\psi^i\right) = \nabla E_i\wedge \psi^i + E_id\psi^i$.

Eventually the collection $(\mathcal{M}, V\mathcal{M},g,\nabla,\varphi)$, where $\mathcal{M}$ is a manifold, $V\mathcal{M}$ is a vector bundle over $\mathcal{M}$ isomorphic to $T\mathcal{M}$, $g$ is a metric on $V\mathcal{M}$, $\nabla$ is a connection on $V\mathcal{M}$ which respects $g$ and $\varphi$ is a 1-form on $\mathcal{M}$ with values in $V\mathcal{M}$ forms the natural data for the Palatini (or the Ashtekar, depending on the gauge group) formulation of gravitation, where the Euler--Lagrange system (\ref{einstein}) is replaced by the system
\begin{equation}\label{palatini}
\left\{ \begin{array}{ccl}
d^\nabla \varphi & = & 0 \\
\lambda_\ell:= \epsilon_{ijk\ell}\Omega^{ij}\wedge \varphi^k & = & 0,
\end{array}\right.
\end{equation}
where $\epsilon_{ijk\ell}$ is the completely skewsymmetric tensor such that $\epsilon_{1234} = 1$, $\Omega^{ij}:= \Omega^i_kg^{kj}$ and $\Omega^i_j:= d\omega^i_j + \omega^i_k\wedge \omega^k_j$ is the curvature 2-form of the connection $\nabla$. Note that (\ref{palatini}) is the Euler--Lagrange equation of the Palatini action $\mathcal{P}[\nabla, \varphi] := \int_\mathcal{M}\epsilon_{ijk\ell}\Omega^{ij}\wedge \varphi^k\wedge \varphi^\ell$. A variant of this formulation is the Ashtekar action. As we have seen the first equation in (\ref{palatini}) is a compatibility condition between $\nabla$ and $\varphi$, the torsion free condition, whereas, once we know that $\varphi^*\nabla$ is torsion free, the second equation of (\ref{palatini}) reads $R_{ij} - {1\over 2}R(\varphi^*g)_{ij} = 0$, i.e. the Einstein equation (\ref{einstein}) in the vacuum.

\subsection{Two ways to understand the solder form}

The standard interpretation of the name `solder form' is the following: through the isomorphism $\varphi_\textsc{m}$, \emph{the solder form glues each fiber $V_\textsc{m}\mathcal{M}$ to the tangent space $T_\textsc{m}\mathcal{M}$ and hence to $\mathcal{M}$}. Then if furthermore $V\mathcal{M}$ is equipped with a connection, telling us how to transport in a parallel way a vector in the fiber $V_\textsc{m}\mathcal{M}$ to an infinitesimally close other fiber, we also obtain a connection on $T\mathcal{M}$. The torsion free condition (\ref{torsionfree1}) then means that we require that the Lie bracket of vector fields on $\mathcal{M}$  agrees with the commutators of infinitesimal parallel transports through $\nabla$. So the manifold $\mathcal{M}$, which has almost no structure (beside the differential structure) without the soldering, acquires with the soldering a much more rich and rigid structure.

Recently I realized that there is an alternative way to understand the name `solder form': instead of being a way to glue the fibers $V_\textsc{m}\mathcal{M}$ to $\mathcal{M}$, \emph{the solder form, together with the connection, allows to solder \textbf{together} the fibers of $V\mathcal{M}$}. In imaged terms one could say that, instead of using $\mathcal{M}$ as a supporting elastic shape on which we glue the fibers $V_\textsc{m}\mathcal{M}$, one uses it as fluid where the $V_\textsc{m}\mathcal{M}$'s are floating and we try to sew together these $V_\textsc{m}\mathcal{M}$'s. Alternatively the points in the geometry we are interested in are not the points of $\mathcal{M}$ but the origins of fibers $V_\textsc{m}\mathcal{M}$. Hence the main object is not $\mathcal{M}$ but the collection of all fibers $\left(V_\textsc{m}\mathcal{M}\right)_{\textsc{m}\in \mathcal{M}}$, soldered together by the connection $\nabla$ and the solder form $\varphi$. In this vision, fibers $V_\textsc{m}\mathcal{M}$ are still the rigid objects and the resulting Riemannian manifold inherits his rigidity from these fibers.
I believe that this point of view is closer to the intuition that the equation of general relativity is an equation on space-time itself and helps to answer the question raised in paragraph \ref{motivation}.

In the following we assume the second point of view and describe the soldering process in more details. In an intuitive manner it may be decomposed into the following steps:
\begin{itemize}
\item  each fiber $V_\textsc{m}\mathcal{M}$ represents the vector space of all infinitesimal displacements of its origin;
\item to each vector $v\in V_\textsc{m}\mathcal{M}$, we can associate the vector $\xi\in T_\textsc{m}\mathcal{M}$ such that $\varphi_\textsc{m}(\xi) = v$, so that, letting $\varepsilon$ be an infinitesimal parameter, we can associate to the end point of $\varepsilon v$ the infinitesimally close fiber $V_{\textsc{m}+\varepsilon\xi}\mathcal{M}$;
\item we glue the origin of $V_{\textsc{m}+\varepsilon\xi}\mathcal{M}$ to the end point of $\varepsilon v$;
\item we transport in a parallel way all vectors in $V_\textsc{m}\mathcal{M}$ to vectors in $V_{\textsc{m}+\varepsilon\xi}\mathcal{M}$, by using the connection $\nabla$.
\end{itemize}
Then condition (\ref{torsionfree1}) can be interpreted by the property that, given $v_1,v_2\in V_\textsc{m}\mathcal{M}$, if we perform the previous process by an infinitesimal displacement $\varepsilon v_1$ first, followed by an infinitesimal displacement by $\varepsilon$ time the paralell transport of $v_2$ along $\varepsilon v_1$, we reach the same point as the point reached by a similar process where the roles of $v_1$ and $v_2$ are exchanged. This is more or less the content of relation (\ref{identite}).

We have up to now described an infinitesimal process, but we need to understand finite analogues of it: this requires an `integration' process (in the general sense of integrating a differential equation), which we call in the following a \emph{solder-integration}.

Note that the key ingredients in the preceding construction are the vector bundle $V\mathcal{M}$, its connection $\nabla$ and the solder form $\varphi$, however the metric $g$ is not essential in this picture (unless we are interested in producing a Riemannian manifold).

\section{Towards more general geometries}\label{pegal1}

Now our `second' point of view suggests natural generalizations. First nothing forces us to suppose that the elementary pieces $V_\textsc{m}\mathcal{M}$ are isomorphic to the tangent spaces $T_\textsc{m}\mathcal{M}$: one could just assume that they are vector spaces. We call the following data a\footnote{We call it `puzzle' because of the intuitive idea is that $(\mathcal{M},V\mathcal{M},\nabla,\varphi)$ represents an infinitesimal puzzle, whose integration is supposed to give a manifold. The `0-' means that we solder together points, which are 0-dimensional.} \textbf{0-puzzle}:
\[
(\mathcal{M},V\mathcal{M},\nabla,\varphi),
\]
where $V\mathcal{M}$ is a vector bundle over $\mathcal{M}$, $\nabla$ is a connection on $V\mathcal{M}$ and $\varphi\in V\mathcal{M}\otimes \Omega^1(\mathcal{M})$ is a 1-form with values in $V\mathcal{M}$. We will say that the 0-puzzle $(\mathcal{M},V\mathcal{M},\nabla,\varphi)$ is \textbf{integrable} if it satisfies the equation
\begin{equation}\label{main}
d^\nabla \varphi = 0.
\end{equation}
We are then interested in the geometry obtained by `solder-integrating' a 0-puzzle. But we do not assume that the rank $n$ of $V\mathcal{M}$ (i.e. the dimension of the fibers $V_\textsc{m}\mathcal{M}$) is the same as the dimension $m$ of $\mathcal{M}$ in general. In the following we consider simple cases where the rank of $\varphi$ is constant. In the case where $V\mathcal{M}$ is equipped with a metric $g$ and $\nabla$ respects $g$, we call the data $(\mathcal{M},V\mathcal{M},g,\nabla,\varphi)$ a \textbf{Riemannian 0-puzzle}.

A useful notion is the following: if $(\mathcal{N},V\mathcal{N},\nabla,\varphi)$ is a 0-puzzle, if $\mathcal{M}$ is another manifold and $u:\mathcal{M}\longrightarrow \mathcal{N}$ is a smooth map, then we can pull-back the bundle $V\mathcal{N}$, the connection $\nabla$ and the 1-form $\varphi$ by $u$ to obtain respectively the bundle $u^*V\mathcal{N}$, the connection $u^*\nabla$ and the 1-form $u^*\varphi$ over $\mathcal{M}$. We then say that \textbf{$(\mathcal{M},u^*V\mathcal{N},u^*\nabla,u^*\varphi)$ is the pull-back by $u$ of $(\mathcal{N},V\mathcal{N},\nabla,\varphi)$}. It easy to check that, \emph{if $(\mathcal{N},V\mathcal{N},\nabla,\varphi)$ is integrable, then $(\mathcal{M},u^*V\mathcal{N},u^*\nabla,u^*\varphi)$ is integrable.}

\subsection{$m=n$ and $\varphi$ is an isomorphism}
This corresponds to the `classical' solder form for a vector bundle already discussed in the previous section. It is well-known that the obtained geometry depends strongly on the holonomy group $\mathfrak{G}$ of the connection $\nabla$, which is a subgroup of $GL(n,\R)$. Then the connection $\varphi^*\nabla$ induced by the solder form has the same holonomy group and hence $T\mathcal{M}$ acquires a $\mathfrak{G}$-structure (see \cite{chern}). In the case where $\mathfrak{G}$ is $O(n)$ we recover the Riemannian geometry.

\subsection{$m<n$ and $\varphi$ is injective}\label{injectivecase}
It means that each $\varphi_\textsc{m}$ embedds $T_\textsc{m}\mathcal{M}$ in $V_\textsc{m}\mathcal{M}$. Actually this case was considered by Ehresmann in \cite{ehresmann}, page 44, where it is connected to a so-called \emph{structure de Cartan au sens large}. In the following we set $H_\textsc{m}\mathcal{M}:= \varphi_\textsc{m}(T_\textsc{m}\mathcal{M})$ and denote by $H\mathcal{M}$ the corresponding bundle. Then we are led to consider the family of subspaces $H_\textsc{m}\mathcal{M}\subset V_\textsc{m}\mathcal{M}$ in a way similar to a distribution of subspaces in some manifolds. Recall that such a distribution is associated to a Pfaffian system and, thanks to Frobenius' theorem, is locally tangent to a foliation by submanifolds iff it satisfies some integrability conditions. Similarly we would like to solder-integrate these subspaces together and here enter the connection $\nabla$ and the solder form $\varphi$ into the game and the analogue of the integrability condition will be  relation (\ref{main}). However the geometric objects that we want to solder-integrate are not only the spaces $H_\textsc{m}\mathcal{M}$'s but also the $V_\textsc{m}\mathcal{M}$'s. But the vectors in $V_\textsc{m}\mathcal{M}$ which are not contained in $H_\textsc{m}\mathcal{M}$ cannot be integrated, i.e. are not tangent to an extended object. Indeed the directions in $V_\textsc{m}\mathcal{M}/H_\textsc{m}\mathcal{M}$ can be interpreted as a an $(n-m)$-dimensional infinitesimal extra thickness
of the $m$-dimensional manifold obtained by solder-integrating the spaces $H_\textsc{m}\mathcal{M}$.

To get a more precise idea of the geometric object constructed, let us consider the particular case where $V\mathcal{M}$ is endowed with a metric $g$ and $\nabla$ respects $g$. Let $N\mathcal{M}$ be the subbundle of $V\mathcal{M}$ normal to $H\mathcal{M}$, i.e. $\forall \textsc{m}\in \mathcal{M}$, $N_\textsc{m}\mathcal{M}$ is the subspace orthogonal to $H_\textsc{m}\mathcal{M}$ for the metric $g$. In particular we have $\forall \textsc{m}\in \mathcal{M}$, $N_\textsc{m}\mathcal{M}\oplus H_\textsc{m}\mathcal{M} = V_\textsc{m}\mathcal{M}$. The family of orthogonal projections $P^H_\textsc{m}: V_\textsc{m}\mathcal{M}\longrightarrow H_\textsc{m}\mathcal{M}$ associated to this decomposition leads to the definition of the following linear map of fiber bundles:
\[
P^H: V\mathcal{M}\longrightarrow H\mathcal{M}.
\]
Consider the connection $\nabla^H$ acting on the set $\Gamma(\mathcal{M},H\mathcal{M})$ of sections of $H\mathcal{M}$  defined by
\[
\forall \sigma \in \Gamma(\mathcal{M},H\mathcal{M}),\quad
\nabla^H\sigma:= P^H(\nabla \sigma).
\]
Then this connection respects the metric induced by $g$ on $H\mathcal{M}$. Moreover we claim that the data $(\mathcal{M},H\mathcal{M},\varphi,\nabla^H)$ (where we consider $\varphi$ as a fiber bundle isomorphism from $T\mathcal{M}$ to $H\mathcal{M}$) is integrable and its integration leads to the Riemannian manifold $(\mathcal{M},\varphi^*g)$. Indeed let $\mathcal{O}\subset \mathcal{M}$ be some open subset such that there exists a moving frame $(e_1,\cdots ,e_n)$ of $V\mathcal{M}$ over $\mathcal{O}$. We assume without loss of generality that $(e_1,\cdots,e_m)$ is a frame of $H\mathcal{M}$ whereas $(e_{m+1},\cdots,e_n)$ is a frame of $N\mathcal{M}$. We denote by $a,b,\cdots$ the indices running from $1$ to $m$, by $\mu,\nu,\cdots$ the indices from $m+1$ to $n$ and by $i,j,\cdots$ the totality of the indices.  We denote by $\omega^i_j$ the connection 1-forms of the connection $\nabla$ on $V\mathcal{M}$: any section $\sigma$ of $V\mathcal{M}$ can be written $\sigma = e_i\sigma^i$ and its covariant derivative reads $\nabla \sigma = e_i\left(d\sigma^i +  \omega^i_j\sigma^j\right)$. Hence in particular if $\sigma$ is a section of $H\mathcal{M}$, then it has the decomposition $\sigma = e_a\sigma^a$ and then $\nabla \sigma = e_a\left(d\sigma^a +  \omega^a_b\sigma^b\right) + e_\mu \omega^\mu_b\sigma^b$ whereas $\nabla^H\sigma = e_a\left(d\sigma^a +  \omega^a_b\sigma^b\right)$.

Now the 1-form $\varphi$ has the representation $\varphi = e_a\varphi^a$ and condition (\ref{main}) reads:
\begin{equation}\label{torsionfree2}
0 = d^\nabla \varphi = e_a\left(d\varphi^a + \omega^a_b\wedge \varphi^b\right) + e_\mu\left(\omega^\mu_b\wedge \varphi^b\right).
\end{equation}
We remark that it implies that
\[
d^{\nabla^H} \varphi = e_a\left(d\varphi^a + \omega^a_b\wedge \varphi^b\right) = 0,
\]
proving that the claim that $(\mathcal{M},H\mathcal{M},\varphi,\nabla^H)$ is integrable. On the other hand the remaining relations $\omega^\mu_b\wedge \varphi^b = 0$, $\forall \mu$ imply that there exists a family of smooth coefficients $\left(h_{\mu ab}\right)_{\mu,a,b}$ such that
\begin{equation}\label{secondff}
\omega^\mu_a = h_{\mu ab}\varphi^b,
\end{equation}
and which satisfies the symmetry condition
\begin{equation}\label{symsecondff}
h_{\mu ab} = h_{\mu ba}.
\end{equation}
The extra data $\left(h_{\mu ba}\right)_{\mu ba}$ can be interpreted as a second fundamental form of some isometric embedding of $(\mathcal{M},\varphi^*g)$ in a Riemannian manifold of dimention $m+n$.

We can actually construct an example of such an embedding, in the total space of $N\mathcal{M}$. For that purpose it will be useful to denote by $A^\mu_{a\nu}$ the coefficients such that
\begin{equation}\label{Amunu}
\omega^\mu_\nu = A^\mu_{a\nu}\varphi^a.
\end{equation}
We let
\[
\Pi: N\mathcal{M}\longrightarrow \mathcal{M}
\]
be the canonical projection of the bundle $N\mathcal{M}$. We identify $\mathcal{M}$ with the zero section of $N\mathcal{M}$. It will be convenient to make further assumptions on the moving frame $(e_1,\cdots,e_n)$ of $V\mathcal{M}$ over $\mathcal{O}\subset \mathcal{M}$: for the first $m$ vectors, we suppose that we are given a local chart $x:\mathcal{O}\longrightarrow \R^m$ and that $e_a:= \varphi_*\left({\partial \over \partial x^a}\right)$, for $1\leq a\leq m$, for the last $n-m$ vectors, we assume that $(e_{m+1},\cdots,e_n)$ is an \emph{orthonormal} frame of the bundle $N\mathcal{M}$. Then for any $\textsc{m}\in \mathcal{O}$ and $\forall y\in N_\textsc{m}\mathcal{M}$, we let $t= (t^{m+1},\cdots,t^n)$ such that $y = e_\mu(\textsc{m})t^\mu$. Hence, denoting by $N_\mathcal{O}\mathcal{M}:= \Pi^{-1}(\mathcal{O})$, we obtain a local chart $(x,t): N_\mathcal{O}\mathcal{M}\longrightarrow x(\mathcal{O})\times \R^{n-m}$, $(\textsc{m},y)\longmapsto (x(\textsc{m}),t)$. Using these coordinates we define the symmetric bilinear form $G = G_{ab}dx^a\otimes dx^b + G_{a\mu}dx^a\otimes dt^\mu + G_{\mu a}dt^\mu \otimes dx^a + G_{\mu\nu}dt^\mu \otimes dt^\nu$ on $N_\mathcal{O}\mathcal{M}$ by:
\[
\begin{array}{l}
G_{ab}(x,t):= g_{ab}(x) - 2t^\mu h_{\mu ab}(x),\quad
G_{\mu\nu}(x,t):= \delta_{\mu\nu},\\ G_{a\mu}(x,t) = G_{\mu a}(x,t):= t^\nu\left(A^\mu_{a\nu}(x) + g_{ab}(x)S^b_{\mu\nu}(x)\right),
\end{array}
\]
where the coefficients $S^b_{\mu\nu}$ can be chosen arbitrarily provided they obey the symmetry condition $S^b_{\mu\nu} = S^b_{\nu\mu}$. Note moreover that, since $(e_{m+1},\linebreak \cdots,e_n)$ is orthonormal, the coefficients $A^\mu_{a\nu}$ defined by (\ref{Amunu}) satisfy $A^\mu_{a\nu} + A^\nu_{a\mu} = 0$. Then, for $|t|$ sufficiently small, $G_{ab}$ is positive definite and hence defines a Riemannian metric on a neighbourhood of $\mathcal{O}$ in $N_\mathcal{O}\mathcal{M}$. Now consider the Levi-Civita connection $\nabla^G$ for this Riemannian manifold and denote by $\varpi^i_j$ its connection forms in the moving frame\linebreak $\left({\partial \over \partial x^1},\cdots,{\partial \over \partial x^m},{\partial \over \partial t^{m+1}},\cdots,{\partial \over \partial t^n}\right)$. One can then compute the value of $\varpi^i_j$ on $\mathcal{M}$ (i.e. for $t = 0$):
\[
\begin{array}{ccl}
\varpi^a_b & = & \omega^a_b - g^{ac}h_{\mu bc}dt^\mu + O(t)\\
\varpi^a_\mu & = & \omega^a_\mu + S^a_{\mu\nu}dt^\nu + O(t)\\
\varpi^\mu_a & = & \omega^\mu_a + A^\mu_{a\nu}dt^\nu + O(t)\\
\varpi^\mu_\nu & = & \omega^\mu_\nu.
\end{array}
\]
It follows that, in particular, if $\iota:\mathcal{M}\hookrightarrow N\mathcal{M}$ denotes the embedding map, $\iota^*\varpi^i_j = \omega^i_j$. Hence the pull-back connection $\iota^*\nabla^G$ coincides with $\varphi^*\nabla$. Similarly $\varphi$ is the pull-back by $\iota$ of the solder form $Id_{T(N\mathcal{M})}$ on $N\mathcal{M}$. In conclusion:
\begin{prop}\label{proposition1}
If $(\mathcal{M},V\mathcal{M},g,\nabla,\varphi)$ is an integrable Riemannian 0-puzzle such that $\varphi$ is injective everywhere, then, up to the restriction to a sufficiently small open subset $\mathcal{O}\subset \mathcal{M}$, there exists an isometric embedding $\iota:(\mathcal{M},g)\longrightarrow (N\mathcal{M},G)$ such that $(\mathcal{M},V\mathcal{M},g,\nabla,\varphi)$ is the pull-back by $\iota$ of $(N\mathcal{M},T(N\mathcal{M}),G,\nabla^G,Id_{T(N\mathcal{M})})$.
\end{prop}

\subsection{$m>n$ and $\varphi$ is surjective}\label{surjectivecase}

In this case, at each point $\textsc{m}\in \mathcal{M}$, $\varphi_\textsc{m}$ has a non trivial kernel, that we will denote by $K_\textsc{m}$. We thus obtain a distribution $K:= \left( K_\textsc{m}\right)_{\textsc{m}\in \mathcal{M}}$ of vector subspaces of dimension $m-n$. Note that condition (\ref{main}), which can be written in a moving frame:
\[
d\varphi^i = - \omega^i_j\wedge \varphi^j,\quad \forall i= 1,\cdots ,n,
\]
reads that the Pfaffian system $\varphi^1 = \cdots = \varphi^n = 0$ satisfies the integrability hypothesis of Frobenius' theorem. Hence there exists an $(m-n)$-dimensional foliation whose leaves are the integral manifolds of the distribution $K$. Following our point of view  a `point' in the geometry we want to build corrresponds to a leaf $\Sigma$ of this foliation. Moreover each leaf $\Sigma$ is equipped with the pull-back bundle $V\Sigma:= \iota_\Sigma^*V\mathcal{M}$ of the immersion map $\iota_\Sigma: \Sigma \hookrightarrow \mathcal{M}$ and $V\Sigma$ can be seen as the tangent space at $\Sigma$ to our geometry. Indeed if $\Sigma'$ is another leaf which is infinitesimally close to $\Sigma$, we can think that $V\Sigma'$ is soldered to $V\Sigma$ along $\Sigma$ as follows. If $\textsc{m}\in \Sigma$ and if $\xi\in T_\textsc{m}\mathcal{M}$ is such that, for an infinitesimal $\varepsilon$, the end point of $\varepsilon\xi$ is in $\Sigma'$, then we solder the origin of the fiber at this end point with $\varepsilon\varphi_\textsc{m}(\xi)$ (observe that $\varphi_\textsc{m}(\xi)$ does not depend on the choice of $\xi$, but on the leaf $\Sigma'$ which contains the end point of $\varepsilon\xi$).

Actually we can be more precise and show the following.
\begin{lemm}\label{propositionlemma}
Let $\mathcal{M}$ be an $m$-dimensional manifold, $V\mathcal{M}$ a rank $n$ vector bundle over $\mathcal{M}$ (where $m>n$), $\nabla$ a connection on $V\mathcal{M}$ and $\varphi$ a 1-form on $\mathcal{M}$ with values in $V\mathcal{M}$ which satisfies (\ref{main}) and with a maximal rank $n$. Then
\begin{enumerate}
\item For any smooth immersion
\[
\begin{array}{cccc}
\gamma:& [-1,1]\times [-1,1] & \longrightarrow & \mathcal{M}\\
& (t,s) & \longmapsto & \gamma(t,s) = \gamma_s(t)
\end{array}
\]
such that the image of $\gamma_0$ is contained in some leaf $\Sigma_0$, \!$\nabla_{\partial \gamma\over \partial t} (\varphi_* {\partial u\over \partial s}) \!= 0$, $\forall (t,s)\in [-1,1]^2$, iff the image of each curve $\gamma_s$ is contained in some leaf $\Sigma_s$;
\item for each leave $\Sigma$, $\iota_\Sigma^*\nabla$, the pull-back by $\iota_\Sigma$ of the connection $\nabla$, is locally flat.
\end{enumerate}
\end{lemm}
\emph{Proof} --- We first show an identity satisfied by any immersion $\gamma: [-1,1]^2\longrightarrow \mathcal{M}$. We start by writing Equation (\ref{main}) using a moving frame $(e_1,\cdots, e_n)$ on $V\mathcal{M}$: $d\varphi^i + \omega^i_j\wedge \varphi^j = 0$. This implies $d(\gamma^*\varphi^i) + (\gamma^*\omega^i_j)\wedge (\gamma^*\varphi^j) = 0$ and hence

\begin{eqnarray}
-d(\gamma^*\varphi^i)\left({\partial \over \partial t},{\partial \over \partial s}\right) &=& (\gamma^*\omega^i_j)\wedge (\gamma^*\varphi^j)\left({\partial \over \partial t},{\partial \over \partial s}\right) \nonumber\\
&=& \omega^i_j\left({\partial \gamma\over \partial t}\right)\varphi^j\left({\partial \gamma\over \partial s}\right) - \omega^i_j\left({\partial \gamma\over \partial s}\right)\varphi^j\left({\partial \gamma\over \partial t}\right).\nonumber
\end{eqnarray}
On the other hand a formula of Cartan gives us:
\begin{eqnarray}
d(\gamma^*\varphi^i)\left({\partial \over \partial t},{\partial \over \partial s}\right) &=&{\partial \over \partial t}\left((\gamma^*\varphi^i)\left({\partial \over \partial s}\right)\right)\nonumber\\
& & - {\partial \over \partial s}\left((\gamma^*\varphi^i)\left({\partial \over \partial t}\right)\right)- (\gamma^*\varphi^i)\left(\left[{\partial \over \partial t},{\partial \over \partial s}\right]\right).\nonumber
\end{eqnarray}
Hence a comparison between both relations leads to:\small
\begin{equation}\label{identite}
{\partial \over \partial t}\!\left(\!\varphi^i\left({\partial \gamma\over \partial s}\right)\!\right) \!+ \omega^i_j\left(\!{\partial \gamma\over \partial t}\!\right)\varphi^j\left({\partial \gamma\over \partial s}\right) =
{\partial \over \partial s}\!\left(\!\varphi^i\!\left({\partial \gamma\over \partial t}\right)\!\right) \!+ \omega^i_j\left(\!{\partial \gamma\over \partial s}\!\right)\!\varphi^j\!\left(\!{\partial \gamma\over \partial t}\right)\!.
\end{equation}\normalsize

We can now prove (i). The condition $\nabla_{\partial \gamma\over \partial t} (\varphi_* {\partial \gamma\over \partial s}) = 0$ is satisfied iff the left hand side of (\ref{identite}) vanishes, and so, iff the right hand side of (\ref{identite}) vanishes also, i.e.
\begin{equation}\label{equadiff}
{\partial \over \partial s}\left(\varphi^i\left({\partial \gamma\over \partial t}\right)\right) = - \omega^i_j\left({\partial \gamma\over \partial s}\right)\varphi^j\left({\partial \gamma\over \partial t}\right).
\end{equation}
This is a homogeneous linear ordinary differential system in $\varphi^i\left({\partial \gamma\over \partial t}\right)$. Now the fact that $\gamma_0$ lies in some leaf $\Sigma_0$ is equivalent to the fact that ${\partial \gamma\over \partial t}(t,0)$ is contained in $K_{\gamma(t,0)}$, the kernel of $\varphi_{\gamma(t,0)}$ and, hence that $\varphi^i\left({\partial \gamma\over \partial t}\right)$ vanishes for $s = 0$. This can be used as an initial condition in system (\ref{equadiff}). Hence we deduce that $\nabla_{\partial \gamma\over \partial t} (\varphi_* {\partial \gamma\over \partial s}) = 0$ iff $\varphi^i\left({\partial \gamma\over \partial t}\right)$ vanishes everywhere, which means that each $\gamma_s$ is contained in some leaf.

Now let us prove (ii). Let $B^{m-n}$ and $B^n$ the unit balls in $\R^{m-n}$ and $\R^n$ respectively and denote by $\vec{t} = (t^1,\cdots ,t^{m-n})$ and $\vec{s} = (s^1,\cdots,s^n)$ the points in $B^{m-n}$ and $B^n$ respectively. Let
\[
\begin{array}{cccc}
X: & B^{m-n}\times B^n & \longrightarrow & \mathcal{M}\\
& (\vec{t},\vec{s}) & \longmapsto & X(\vec{t},\vec{s})
\end{array}
\]
be a local parametrization such that, $\forall \vec{s}\in B^n$, 
\[
\begin{array}{cccc}
X_{\vec{s}}:& B^{m-n} & \longrightarrow & \mathcal{M}\\
& \vec{t} & \longmapsto & X(\vec{t},\vec{s})
\end{array}
\]
is a local parametrization of a leaf, which we denote by $\Sigma_{\vec{s}}$. This is equivalent to the condition that $\varphi_X({\partial X\over \partial t^\mu}) = 0$, $\forall \mu = 1,\cdots, m-n$.

We first observe that $\left(\varphi\left({\partial X\over \partial s^1}\right), \cdots ,\varphi\left({\partial X\over \partial s^n}\right)\right)$ is a moving frame of $V\mathcal{M}$ over $X([-1,1]^2)$. Now fix $\mu$ and $a$ such that $1\leq \mu\leq m-n$ and $1\leq a\leq n$ and apply (\ref{identite}) with $\gamma(t,s) = X(\vec{t},\vec{s})|_{t^\mu=t,s^a=s}$ to obtain
\[
\nabla_{\partial X\over \partial t^\mu}\left( \varphi\left({\partial X\over \partial s^a}\right)\right) = 
\nabla_{\partial X\over \partial s^a}\left( \varphi\left({\partial X\over \partial t^\mu}\right)\right).
\]
But since the right hand side of this relation vanishes because of $\varphi({\partial X\over \partial t^\mu}) = 0$, we conclude that $\varphi\left({\partial X\over \partial s^a}\right)$ is parallel for $\nabla$ along $\Sigma_{\vec{s}}$. We hence conclude that each restriction bundle $V\Sigma_{\vec{s}}$ can be locally trivialized by using the parallel frame $\left(\varphi\left({\partial X\over \partial s^1}\right), \cdots ,\varphi\left({\partial X\over \partial s^n}\right)\right)$. This proves (ii).\hfill $\square$\\

Note that claim (i) in Proposition \ref{propositionlemma} means that a property of paralellism between the leaves $\Sigma$ is satisfied. A consequence of (ii) in Proposition \ref{propositionlemma} is that, for each leaf $\Sigma$, the bundle $V\Sigma$ with the connection $\iota_\Sigma^*\nabla$ can be identified with $\Sigma \times \R^n$ with the flat connection. Hence, if we denote by $\mathcal{Q}$ the set of leaves which intersect the set $\mathcal{O}:= X(B^{m-n}\times B^n)\subset \mathcal{M}$, we can construct the rank $n$ vector bundle $V\mathcal{Q}$ over $\mathcal{Q}$ whose fiber over each leaf $\Sigma\cap \mathcal{O}$ is the set of sections of $V\Sigma$ which are parallel for $\iota_\Sigma^*\nabla$ and endow this bundle with the connection $\overline{\nabla}$ induced by $\nabla$. Then, if $Q:\mathcal{O}\longrightarrow \mathcal{Q}$ is the quotient map, we can conclude that, over $\mathcal{O}$, $V\mathcal{M}$ is the pull-back bundle by $Q$ of $V\mathcal{Q}$ and that, similarly, $\nabla = Q^*\overline{\nabla}$. Moreover $\varphi$ is the pull-back of a covariantly closed 1-form $\overline{\varphi}$ on $\mathcal{Q}$ with values in $V\mathcal{Q}$ which is an isomorphism. In particular, if $V\mathcal{M}$ is endowed with a metric and if $\nabla$ is compatible with this metric, the fibers are equi-distant, which implies that $\mathcal{Q}$ is endowed with a Riemannian metric. Hence:

\begin{prop}\label{proposition2}
If $(\mathcal{M},V\mathcal{M},\nabla,\varphi)$ is an integrable 0-puzzle such that $\varphi$ is surjective everywhere, then, up to the restriction to a sufficiently small open subset $\mathcal{O}\subset \mathcal{M}$, there exists an integrable 0-puzzle $(\mathcal{Q},V\mathcal{Q},\overline{\nabla},\linebreak \overline{\varphi})$ and a submersion $Q:\mathcal{M}\longrightarrow \mathcal{Q}$ such that $(\mathcal{M},V\mathcal{M},\nabla,\varphi)$ is the pull-back by $Q$ of $(\mathcal{Q},V\mathcal{Q},\overline{\nabla},\overline{\varphi})$. If furthermore $(\mathcal{M},V\mathcal{M},g,\nabla,\varphi)$ is Riemannian, then $(\mathcal{Q},V\mathcal{Q},\overline{g},\overline{\nabla},\overline{\varphi})$ is so and $g=Q^*\overline{g}$.
\end{prop}

\subsubsection{The general case where $\varphi$ has a constant rank}\label{general}
Assume that $\mathcal{M}$ is $m$-dimensional, $V\mathcal{M}$ is of rank $n$ and $\varphi$ has a constant rank equal to $k$, such that $k\leq \hbox{inf}(m,n)$. As in paragraph \ref{surjectivecase}, we set $K_\textsc{m}:= Ker\varphi_\textsc{m}$ and we denote by $K = (K_\textsc{m})_{\textsc{m}\in \mathcal{M}}$ the corresponding distribution. Then again $K$ is integrable and its integration provides a foliation by leaves $\Sigma$ of dimension $m-k$. One can locally factorize $V\mathcal{M}$ and its connection $\nabla$ by a quotient map $Q:\mathcal{O}\longrightarrow \mathcal{Q}$ where $\mathcal{Q}$ is the set of leaves of $K$ intersecting some open subset $\mathcal{O}\subset \mathcal{M}$:  $V\mathcal{M}$, $\nabla$ and $\varphi$ are locally the pull-back by $Q$ of respectively a bundle $V\mathcal{Q}$, a connection $\overline{\nabla}$ over $V\mathcal{Q}$ and a 1-form $\overline{\varphi}$ on $\mathcal{Q}$ with values in $V\mathcal{Q}$ which is covariantly closed.

In the case where $(\mathcal{M},V\mathcal{M},g,\nabla,\varphi)$ is Riemannian, we can further apply Proposition \ref{proposition1} to the 0-puzzle $(\mathcal{Q},V\mathcal{Q},\overline{g},\overline{\nabla},\overline{\varphi})$ and locally embedd isometrically $(\mathcal{Q},\overline{g})$ in an $n$-dimensional Riemannian manifold $(N\mathcal{Q},G)$ with its Levi-Civita connection $\nabla^G$ and its solder form $Id_{T(N\mathcal{Q})}$. Hence:
\begin{prop}\label{proposition3}
If $(\mathcal{M},V\mathcal{M},g,\nabla,\varphi)$ is an integrable Riemannian 0-puzzle such that $\varphi$ has a constant rank, then, up to the restriction to a sufficiently small open subset $\mathcal{O}\subset \mathcal{M}$, there exists an $n$-dimensional manifold $(N\mathcal{Q},G)$ and a smooth map $u$ ($:= \iota\circ Q$, where $\iota: (Q,\overline{g})\longrightarrow (N\mathcal{Q},G)$ is an isometric embedding as in paragraph \ref{injectivecase}) such that $(\mathcal{M},\linebreak V\mathcal{M},g,\nabla,\varphi)$ is the pull-back by $u$ of $(N\mathcal{Q},T(N\mathcal{Q}),G,\nabla^G,Id_{T(N\mathcal{Q})})$.
\end{prop}

\section{More general geometries}
A further generalization consists in replacing the solder 1-form by a $p$-form, where $1\leq p\leq m$. We call a \textbf{$(p-1)$-puzzle} a data:
\[
(\mathcal{M},V\mathcal{M},\nabla,\varphi),
\]
where, as before, $\mathcal{M}$, $V\mathcal{M}$ and $\nabla$ are respectively a manifold, a vector bundle and a connection, but, in contrast, $\varphi$ is a $p$-form with coefficients in $V\mathcal{M}$ (i.e. a section of $V\mathcal{M}\otimes_\mathcal{M} \Omega^p(\mathcal{M})$). We say that this \textbf{$(p-1)$-puzzle is integrable iff $\varphi$ satisfies the condition $d^\nabla \varphi = 0$} which, by using a local moving frame $(e_1,\cdots ,e_n)$ on $V\mathcal{M}$ and using the decomposition $\varphi = e_i\varphi^i$, where $\varphi^i\in \Omega^p(\mathcal{M})$, reads
\[
d\varphi^i + \omega^i_j\wedge \varphi^j = 0.
\]
In the following we wish to draw some intuitive picture of a geometry which one could associate to an integrable $(p-1)$-puzzle. Being unable to give a precise definition of this geometry we will just refer at it as `the geometry \emph{upstair}'. We note that, if $p>1$, we do not have anymore a way to solder together two infinitesimally close points, in the sense that there is no privileged way to connect the end point of an infinitesimal vector in a fiber $V_\textsc{m}\mathcal{M}$ with the origin of some other fiber $V_{\textsc{m}'}\mathcal{M}$.\linebreak Actually in the geometry \emph{usptair} $(\mathcal{M},V\mathcal{M},\nabla,\varphi)$ the fundamental objects should be $(p-1)$-dimensional objects rather than points. We do not mean that points do not exist anymore but that submanifolds of dimension less than or equal to $p-1$ have no structure beside the restriction of $V\mathcal{M}$ and its connection over it (exactly like the fact that, in the ordinary geometry, points have no structure, and the various geometric structures---Riemannian, symplectic, etc.---concern the relationship between points). Hence the basic objects will be $(p-1)$-dimensional oriented submanifolds $\Sigma$ of $\mathcal{M}$ equipped with the pull-back bundle $\iota_\Sigma^*V\mathcal{M}$ by the embdedding map $\iota_\Sigma: \Sigma \hookrightarrow \mathcal{M}$.

To understand the intuition behind this, it is useful to go back to the 0-puzzles $(\mathcal{M},V\mathcal{M},\nabla,\varphi)$. Then the 1-form $\varphi$ and the connection $\nabla$ provide informations about how to connect a \emph{pair of points} in the geometry: if $\textsc{m}_0\in \mathcal{M}$ and $\textsc{m}_1\in \mathcal{M}$, then any $\mathcal{C}^1$ path $\gamma:[0,1]\longrightarrow \mathcal{M}$ which connects $\textsc{m}_0$ to $\textsc{m}_1$ can be lifted using $\varphi$ and $\nabla$ and condition (\ref{main}) garantees that all such lifts join a unique pair of points \emph{upstair}. Similarly in the case where $p>1$ we replace paths by $p$-dimensional oriented submanifolds $\Gamma$ with boundary and the pair of points by its boundary $\Sigma:= \partial \Gamma$. Of course $\Sigma$ could then have an arbitrary number of connected components (for instance, if $p=2$, $\Sigma$ could be an emdedded circle) and, more generally, $\Sigma$ could have an arbitrary topology. Then we wish to interpret the condition $d^\nabla \varphi = 0$ as insuring that, if $\Gamma$ and $\Gamma'$ are two oriented $p$-dimensional manifolds of $\mathcal{M}$ connecting the same system of $(p-1)$-dimensional submanifolds, then their lifts \emph{upstair} connect the `same' $(p-1)$-dimensional objects. 

In the following we expound several points of view to illustrate these considerations.

\subsection{The geometry \emph{upstair} as a primitive}

The idea is to think at the geometry \emph{upstair} as a kind of primitive of $\varphi$, in the same way as the fact that, if $\alpha\in \Omega^1(\mathcal{M})$ is a closed 1-form with real coefficients, then, by Poincar{\'e}'s lemma, there exists a real valued function $f$ on $\mathcal{M}$ such that $\alpha = df$. As in the case of Poincar{\'e} lemma, where $f$ is uniquely defined up to a constant and so only the difference $f(\textsc{m}_1) - f(\textsc{m}_0)$ is defined unambiguously, the data $\varphi$ and $\nabla$ just give us a way to relate two points in the geometry \emph{upstair}.

One example given in \cite{helein} helps to understand this idea in the case where $p=1$. It corresponds to a Riemannian 0-puzzle $(\mathcal{M},V\mathcal{M},g,\nabla,\varphi)$. Let $\mathcal{M}$ be a smooth manifold and $(\mathcal{N},g)$ be a Riemannian manifold with its Levi-Civita connection $\nabla^g$. Let $u:\mathcal{M}\longrightarrow \mathcal{N}$ be any smooth map and consider the pull-back bundle $u^*T\mathcal{N}$, i.e. the bundle over $\mathcal{M}$ whose fiber at any $\textsc{m}\in \mathcal{M}$ is $T_{u(\textsc{m})}\mathcal{N}$. We endow $u^*T\mathcal{N}$ with the pull-back connection $u^*\nabla^g$. Then $\varphi:= du$ is a 1-form on $\mathcal{M}$ with values in $u^*T\mathcal{N}$ which satisfies
\[
d^{u^*\nabla^g}\varphi = 0.
\]
This condition can be interpreted as a locally necessary and sufficient condition for $\varphi$ to be the differential of some `nonlinear' 0-form, namely the map $u$ into $\mathcal{N}$. Note that in the case where $\mathcal{M} = \mathcal{N}$ and $\varphi$ is the identity map we recover the solder form for $(\mathcal{M},g)$.

\subsection{Isometric embeddings}

Here we restrict to the case of Riemannian $(p-1$)-puzzles. The \emph{isometric embedding} problem settled in \cite{helein} is the following: given an integrable Riemannian $(p-1)$-puzzle $(\mathcal{M},V\mathcal{M},g,\nabla,\varphi)$, find $N\in \N$ and a vector bundle map $T: V\mathcal{M} \longrightarrow \mathcal{M}\times \R^N$ such that:
\begin{enumerate}
\item $\forall (\textsc{m},v)\in V\mathcal{M}$, $T(\textsc{m},v) = (\textsc{m},T_\textsc{m}(v))$, where, $\forall \textsc{m}\in V\mathcal{M}$, $T_\textsc{m}: (V_\textsc{m}\mathcal{M},g_\textsc{m})\longrightarrow (\R^N,\langle\cdot,\cdot\rangle)$ is an \textbf{isometry};
\item $T_*\varphi$ is \textbf{closed}, i.e. $d(T_*\varphi) = 0$.
\end{enumerate}
Here $T_*\varphi$ is the $p$-form on $\mathcal{M}$ with coefficients in $\R^N$ such that, for any local moving frame $(e_1,\cdots,e_n)$ of $V\mathcal{M}$, if $\varphi = e_i \varphi^i$, with $\varphi^i\in \Omega^j(\mathcal{M})$, then
\[
\left(T_*\varphi\right)_\textsc{m}:= T_\textsc{m}(e_i(\textsc{m})) \varphi^i_\textsc{m}.
\]
We know that this problem has a solution in at least three cases:
\begin{itemize}
\item if $p=1$ and $\varphi$ is an isomorphism, then, without loss of generality, we can assume that $\varphi$ is the identity automorphism $Id_{T\mathcal{M}}$ of $T\mathcal{M}$, for a Riemannian manifold $\mathcal{M}$, and, locally, $T_*\varphi$ is nothing but the exterior differential of an isometric embedding $\Phi:\mathcal{M}\longrightarrow \R^N$ of $(\mathcal{M},g)$ (see \cite{helein});
\item if $p=1$ and $\varphi$ has a constant rank. We can use Proposition \ref{proposition3} which tells us that $(\mathcal{M},V\mathcal{M},g,\nabla,\varphi)$ is the pull-back by some map $u:\mathcal{M}\longrightarrow \mathcal{N}$ of $(\mathcal{N},T\mathcal{N},G,\nabla^G,Id_{T\mathcal{N}})$, where $(\mathcal{N},G)$ is a Riemannian manifold. Then we use an isometric embedding of $(\mathcal{N},G)$ as in the previous case;
\item a local solution, if $p = m-1$ and if the data are real analytic, by the result of N. Kahouadji \cite{kahouadji}. Hence there exists some $\R^N$-valued $(m-2)$-form $\Phi$ on sufficientely small open subsets of $\mathcal{M}$ such that $d\Phi = T_*\varphi$.
\end{itemize}
These results illustrate again the idea that the geometry \emph{upstair} is a `primitive' of $\varphi$ and $\nabla$, i.e. that $\varphi$ is the differential of some non linear $(p-1)$-form, whose precise meaning needs to be defined.

\subsection{Defining observables}
Another way to explore the \emph{upstair} geometry of a $(p-1)$-puzzle is to focus on the set of `functions' on it. It is perhaps safer to use the more generic name of `observables' as in physics, since, for instance in the case where $p>1$, the analogues of functions will be $(p-1)$-forms.

The simplest idea is to define the set of observables on $(\mathcal{M},V\mathcal{M},\linebreak \nabla,\varphi)$ as \emph{the set of $\mathcal{C}^1$ sections $f$ of $V^*\mathcal{M}$, the bundle dual to $V\mathcal{M}$, such that
\begin{equation}\label{dual}
d(f,\varphi) = 0,
\end{equation}
where $(f,\varphi) \in \Omega^p(\mathcal{M})$ is the duality pairing between $f$ and $\varphi$}. Using a moving frame $(e_1,\cdots,e_n)$ on $V\mathcal{M}$ and denoting by $(\eta^1,\cdots,\eta^n)$ its dual frame, we can decompose $\varphi = e_i \varphi^i$ and $f = f_i\eta^i$ and then $(f,\varphi):= f_i\varphi^i$. Equation (\ref{dual}) then reads
\begin{equation}\label{dual1}
\left(df_i - f_j\omega^j_i\right)\wedge \varphi^i = 0.
\end{equation}
Equation (\ref{dual1}) can be difficult to solve in general. For instance, if $p = 1$, using Cartan's lemma, it reduces to solving
\[
df_i - f_j\omega^j_i = h_{ik}\varphi^k,
\]
where the coefficients $h_{ik}$ are continuous functions which satisfy the symmetry condition $h_{ik} = h_{ki}$. There is however another formulation of Equation (\ref{dual}): find \textbf{closed} $p$-forms $\beta\in \Omega^p(\mathcal{M})$ such that there exists $f\in \Gamma(\mathcal{M},V^*\mathcal{M})$ such that $\beta = (f,\varphi)$. Locally there is no loss of generality in assuming that $\beta$ is \emph{exact} and hence in looking for $(p-1)$-forms $\alpha$ on $\mathcal{M}$ such that we have 
\begin{equation}\label{exactform}
d\alpha = (f,\varphi)
\end{equation}
for some $f\in \Gamma(\mathcal{M},V^*\mathcal{M})$. Using (\ref{exactform}) we can hence give a local description of all solutions of (\ref{dual}) for $p=1$ and in the case where the rank of $\varphi$ is constant. For simplicity we assume that $\varphi$ is either injective or surjective as in Section \ref{pegal1}.
\begin{itemize}

\item if $\varphi$ is an isomorphism. This is the simplest case: we start with an arbitrary function $\alpha$, then $d\alpha$ is a 1-form and hence there exists an \textbf{unique} $f\in \Gamma(\mathcal{M},V^*\mathcal{M})$ such that (\ref{exactform}) holds. Indeed using a local moving frame $(e_1,\cdots, e_m)$ on $V\mathcal{M}$ we have the decomposition $\varphi = e_1\varphi^1 + \cdots + e_m\varphi^m$ and the condition that $\varphi$ is an isomorphism means that $(\varphi^1,\cdots,\varphi^m)$ is a local coframe on $\mathcal{M}$. Hence $(f_1,\cdots,f_n)$ are the coefficients of the decomposition of $d\alpha$ in $(\varphi^1,\cdots,\varphi^m)$. So in this case, observables on the geometry upstair coincide with functions on $\mathcal{M}$;

\item if $m<n$ and $\varphi$ is injective. We use the same notations as in paragraph \ref{injectivecase}. Let $(e_1,\cdots,e_n)$ be a moving frame on $V\mathcal{M}$ such that $(e_1,\dots,e_m)$ is a frame of $H\mathcal{M}$, the image of $T\mathcal{M}$ by $\varphi$. Then $(\varphi^1,\cdots,\varphi^m)$ is a coframe. Hence, as in the previous case, for any function $\alpha$ on $\mathcal{M}$, there exist unique coefficients $f_1,\cdots ,f_m$ such that $d\alpha = f_1\varphi^1 + \cdots + f_m\varphi^m = (\underline{f},\varphi)$, where $\underline{f}:= (f_1,\cdots,f_m,0,\cdots ,0)$. However $\underline{f}$ is not the unique section of $V^*\mathcal{M}$ such that $d\alpha = (f,\varphi)$ (and moreover the construction of $\underline{f}$ is not canonical and depends on how we complete $(e_1,\dots,e_m)$ into a basis of $V\mathcal{M}$). Indeed we can add to $\underline{f}$ any section $g$ of $V^*\mathcal{M}$ such that $(g,\varphi) = 0$. In other words, if we define $H^\perp\mathcal{M}:= \{(\textsc{m},\lambda)\in V^*\mathcal{M}|\ \lambda\circ \varphi_\textsc{m} = 0\}$, a vector subbundle of $V^*\mathcal{M}$, then the set of $f\in \Gamma(\mathcal{M},V^*\mathcal{M})$ such that (\ref{exactform}) is $\{ \underline{f} + g|\ g\in \Gamma(\mathcal{M},H^\perp\mathcal{M})\}$. Alternatively the set of smooth $\mathcal{C}^\infty$ observables on the geometry \emph{upstair} coincides with the quotient $\mathcal{C}^\infty(\mathcal{N})/\mathcal{I}^\infty(\mathcal{M})^2$, where $\mathcal{N}$ is an $n$-dimensional manifold in which $\mathcal{M}$ is embedded, $\mathcal{I}^\infty(\mathcal{M})$ is the ideal of $\mathcal{C}^\infty(\mathcal{N})$ composed of functions which vanish on $\mathcal{M}$ and $\mathcal{I}^\infty(\mathcal{M})^2:= \{f^2|\ f\in \mathcal{I}^\infty(\mathcal{M})\}$;

\item if $m>n$ and $\varphi$ is surjective. We use the same notations as in paragraph \ref{surjectivecase}. In that case Equation (\ref{exactform}) may have no solution. Actually this equation admits solutions iff $\alpha$ is constant on each leaf $\Sigma$, i.e. on each integral manifold of the distribution $K:= Ker\varphi$. Hence $\alpha$ corresponds to a function on the set of leaves of the foliation. Of course globally the space of leaves of a foliation has in general a complicated topology and the set of smooth (or even measurable) functions on it may be trivial, unless the foliation is a fibration. An alternative would be to consider this set as a noncommutative space in the sense of Alain Connes.
\end{itemize}

\subsection{Some \emph{upstair} geometries for $p=2$}
We consider a 1-puzzle $(\mathcal{M},V\mathcal{M},\nabla,\varphi)$. First we define, $\forall \textsc{m}\in \mathcal{M}$, the subspace $K_\textsc{m}:= \{\xi\in T_\textsc{m}\mathcal{M}|\ \xi\iN\varphi_\textsc{m} = 0\}$ and the distribution of vector subspaces $K = \left( K_\textsc{m}\right)_{\textsc{m}\in \mathcal{M}}$. We assume that the dimension of $K_\textsc{m}$ is constant and equal to $k$. We claim that this distribution is integrable, i.e. we will prove that if $X,Y$ are smooth vector fields on $\mathcal{M}$ such that $X_\textsc{m},Y_\textsc{m}\in K_\textsc{m}$ everywhere, then $[X,Y]_\textsc{m}\in K_\textsc{m}$ everywhere or, equivalentely, that if $X\iN\varphi = Y\iN\varphi = 0$, then $[X,Y]\iN\varphi = 0$. Let $Z$ be an arbitrary vector field on $\mathcal{M}$, then on the one hand the identity
\[
\begin{array}{ccl}
d\varphi^i(X,Y,Z) & = & X\cdot \varphi^i(Y,Z) + Y\cdot \varphi^i(Z,X) + Z\cdot \varphi^i(X,Y)\\
& & - \varphi^i([X,Y],Z) - \varphi^i([Y,Z],X) - \varphi^i([Z,X],Y)
\end{array}
\]
can be simplified (using $X\iN\varphi = Y\iN\varphi = 0$) as $d\varphi^i(X,Y,Z) =\linebreak - \varphi^i([X,Y],Z)$. On the other hand the identity $\omega^i_j\wedge \varphi^j(X,Y,Z) = \omega^i_j(X)\varphi^j(Y,Z) + \omega^i_j(Y)\varphi^j(Z,X) +\omega^i_j(Z)\varphi^j(X,Y)$ can similarly be simplified as $\omega^i_j\wedge \varphi^j(X,Y,Z) = 0$. Thus, from the relation $0 = d^\nabla(e_i\varphi^i) = e_i(d\varphi^i + \omega^i_j\wedge\varphi^j)$, we deduce:
\[
\varphi^i([X,Y],Z) = -d\varphi^i(X,Y,Z) = \omega^i_j\wedge \varphi^j(X,Y,Z) = 0.
\]
Hence $[X,Y]\iN\varphi = 0$. Thus we can integrate the distribution $K$ into a foliation by leaves of dimension $k$. Then the space of leaves of this foliation can be locally identified with an $(m-k)$-dimensional manifold $\mathcal{Q}$ and, in a way similar to paragraph \ref{surjectivecase}, two curves in $\mathcal{M}$ which are contained in the same leaf of the foliation and are cobordant inside this leaf lift to the same curve \emph{upstair}. 

In the following we assume that the dimension of $\mathcal{M}$ is 4 and we consider very simple examples.
\begin{itemize}
\item $V\mathcal{M}$ is a rank 1 vector bundle and $\varphi$ is decomposable, i.e. of the form $\varphi = \psi\wedge \chi$, where $\psi$ and $\chi$ are 1-forms. Then the distribution $K$ is a distribution of planes. Hence the space of leaves $\mathcal{Q}$ is a surface and two curves in $\mathcal{M}$ lift to the same curve \emph{upstair} iff they are contained in the same leaf of the foliation and they are cobordant inside this leaf. Assuming that $\varphi$ does not vanish, $\varphi$ can be interpreted as a ($V\mathcal{M}$-valued) area 2-form on $\mathcal{Q}$. The observables corresponds then to closed 2-forms on $\mathcal{Q}$, i.e. differentials of 1-forms on $\mathcal{M}$ modulo exact closed 1-forms;

\item $V\mathcal{M}$ is a rank 1 vector bundle and $\varphi$ is symplectic, i.e. of the form $\varphi = e_1\phi$, where $\phi\wedge \phi\neq 0$. Then the distribution $K$ reduces to 0. However a situation similar to the previous case occurs if we replace the leafs of the foliation of the previous case by \emph{Lagrangian} submanifolds: two curves in $\mathcal{M}$ lift to the same curve \emph{upstair} iff they are contained in the same \emph{Lagrangian} submanifold and cobordant inside. Hence the geometry upstair should be connected with the theory of Lagrangian cobordisms initiated by V.I. Arnold \cite{audin};

\item $V\mathcal{M}$ is a rank 2 vector bundle, $\varphi$ has a rank 2 and both components of $\varphi$ are symplectic. Then roughly speaking we need to replace the leafs of the foliation of the first case or the \emph{Lagrangian} submanifolds of the second case by \emph{pseudo-complex} curves. Let us give a very simple example, for $\mathcal{M} = \R^4$ and $V\mathcal{M} = \R^4\times \R^2$ with the flat connection. By identifying $\R^4$ with the quaternions $\H$ we can construct three complex structures, given by the left multiplication by respectively $i$, $j$ and $k$. Let $\langle\cdot,\cdot\rangle$ be the real Euclidean scalar product on $\H$. Then we define $\varphi$ by $\varphi(u,v) = (\langle u,jv\rangle, \langle u,kv\rangle)$. From the identity $u\overline{v} = \langle u,v\rangle + \langle u,iv\rangle i + \langle u,jv\rangle j + \langle u,kv\rangle k$ in $\H$, we easily deduce that any real plane in $\H$ on which $\langle u,jv\rangle$ and $\langle u,kv\rangle$ vanish (i.e. in the kernel of $\varphi$) is a complex plane for the complex structure given by $i$. Hence two curves in $\H$ lift to the same curve \emph{upstair} iff they can be linked by a cobordims by a \emph{complex} surface ;

\item $V\mathcal{M}$ is a rank 3 vector bundle, $\varphi$ has a rank 3 and all components of $\varphi$ are symplectic. This situation occurs if, for instance, $\varphi$ is the curvature of an $SU(2)$-valued self-dual connection. Then each curve usptair lifts exactly one curve in $\mathcal{M}$. However the set of observables on the curves in the geometry \emph{upstair} is different from the set of observables on the set of curves on a 4-dimensional manifold. 
\end{itemize}

\section{Possible physical applications}
We have discussed one important motivation, from the theory of General Relativity, which was at the origin of these speculations. Other motivations or possible applications are discussed in \cite{helein} or in \cite{kahouadji}: many equations in physics take the form $d^\nabla \varphi =0$, meaning that, in a covariant sense, $\varphi$ is an invariant form.

For instance the Bianchi relation for a Yang--Mills connection:
\[
d^\nabla \varphi = 0
\]
is a relation which reflects the fact that $\varphi$ is a covariant derivative of the connection, i.e. $\varphi = F = dA+A\wedge A$, where $A$ is the connection form. The analogous
equation for $*\varphi$ is the Yang--Mills dynamical equation. In general we cannot `integrate' the relation $d^\nabla (*\varphi) = 0$, unless the gauge group is $U(1)$, which corresponds to the classical electromagnetism. This is of course strongly related to our difficulty in extending the electro-magnetic duality to non-linear Yang--Mills theories. Our preceding considerations tend to propose a geometric interpretation to the problem of integrating the equation $d^\nabla (*\varphi) = 0$.

Another example comes from the stress-energy tensor in general relativity which, as explained in \cite{kahouadji}, could be seen as an $(m-1)$-form $\varphi$ with values in the tangent bundle. Again if we could integrate the relation $d^\nabla\varphi = 0$, the resulting \emph{upstair} geometry would encode the mass and momentum and this would provide a way to define them properly on a curved manifold (we note that, in this point of view, it seems artificial to separate the energy and the momentum). 

A last possible area where these ideas may apply could be the theory of (super-)strings (or more generally M-theories). On the one hand if we plan to construct such a theory by respecting the original ideas of general relativity, we should then be able to explain how the space-time is modelled out of some dynamical equations. But on the other hand the string theory requires that the fundamental objects are strings and not points. Hence it is possible that, at a classical (i.e. non quantum) level one needs a geometry \emph{upstair} based on a soldering 2-form. For instance an extremely naive but natural proposal in our framework would be to replace the Einstein--Palatini action by a similar action but involving a connection $\nabla$ and a 2-form $\varphi$.


\end{document}